\documentclass{Odyssey2026}

 \interspeechcameraready

\title{Speaker-Invariant Representation Learning for Spoofing Detection via \\Gradient Reversal and A Variational Information Bottleneck}

\author[affiliation={1}]{Anh-Tuan}{Dao}
\author[affiliation={1}]{Driss}{Matrouf}
\author[affiliation={1}]{Mickael}{Rouvier}
\author[affiliation={2}]{Nicholas}{Evans}

\affiliation{Laboratoire Informatique d’Avignon}{Avignon Universite}{France}
\affiliation{EURECOM }{ Sophia Antipolis}{France }
\email{\{anh-tuan.dao, driss.matrouf, mickael.rouvier\}@univ-avignon.fr, evans@eurecom.fr}


\usepackage{comment}
\usepackage{soul}

\begin{document}

\maketitle

% the abstract here must exactly match the abstract entered into the paper submission system
\begin{abstract}
Sophisticated generative speech technology can undermined the reliability of voice biometrics. While spoofing detection systems excel when assessed under in-domain conditions, generalisation to out-of-domain settings is often poor. In this paper, we show that such issues could be caused by speaker bias, where models learn individual voice traits rather than markers of manipulation or generation.
We propose a teacher-student framework for speaker-invariant spoofing detection that disentangles identity without requiring speaker labels. We leverage a pre-trained speaker recognition teacher to guide a student model via a gradient reversal layer. To control the balance between suppressing cues related to voice identity with the preservation of those related to spoofing detection, we integrate a Variational Information Bottleneck. 
Evaluations across nine datasets show our model achieves a 25.7\% relative reduction to the EER compared to the MHFA baseline.

\end{abstract}

\section{Introduction}
Voice biometric systems remain vulnerable to increasingly sophisticated spoofing attacks generated by modern text-to-speech (TTS) and voice conversion (VC). As a result, spoofing detection systems have become an essential security component to safeguard voice biometric deployments. %Large-scale benchmarks~\cite{ASVspoof24, ADD2022} have played a pivotal role in advancing this field by providing standardized datasets and evaluation protocols, enabling rapid progress in detection performance and robustness.
By providing standardized datasets and evaluation protocols, large-scale benchmarks~\cite{ASVspoof24, ADD2022} have facilitated rapid progress in detection performance and robustness.

Recent years have witnessed a paradigm shift in spoofing detection with the adoption of self-supervised learning (SSL) models~\cite{AASIST,conformer,MHFA_Spoof,rohdin24_but_asvspoof,SLS, Mamba}. By exploiting large-scale pre-training with raw speech, SSL-based models learn expressive representations which substantially outperform conventional supervised architectures~\cite{AASIST1,RawNet2,dao24_asvspoof}. Despite these gains, robust generalization across datasets and attack types remains a challenge. Models trained using one particular corpus often exhibit sharp performance degradations when evaluation is performed with other corpora. %unseen data.

A recognized cause of such fragility is shortcut learning. Rather than focusing on intrinsic bonafide-spoofing artifacts, spoofing detection models may exploit auxiliary correlations embedded in the training data. While the earlier work emphasized low-level shortcuts such as non-speech patterns~\cite{Silence2021}, more recent studies suggest that factors related to higher-level speaker information can also play a  role~\cite{dao2026assessingimpactspeakeridentity}.  TTS systems often generate speech with a canonicalized vocal profile that reflects a characteristic model-voice. If a detector treats these related cues as proxies for the spoofing label, then it can learn to identity the specific model-voice rather than the more generalisable differences between bona fide speech and spoofed speech generated with other techniques or models. This can result in overfitting to the training distribution and poor generalisation.

In our earlier work, we investigated this phenomenon by explicitly suppressing speaker information within spoofing detection models~\cite{dao2026assessingimpactspeakeridentity}. 
% Prior work investigated this phenomenon by explicitly suppressing speaker information within spoofing detection models~\cite{dao2026assessingimpactspeakeridentity}. 
However, directly enforcing speaker invariance via adversarial multi-task learning presents an inherent risk. Aggressive adversarial training may inadvertently discard other correlated cues that are beneficial to spoofing detection. This raises a critical question: how can speaker-related information be removed in a controlled and principled manner? Moreover, the prior approach relied on speaker identity labels within spoofing detection datasets, which are not consistently available across corpora. Even when such labels exist, the limited number and diversity of speakers in many spoofing datasets may be insufficient to learn speaker characteristics that generalize beyond the training domain.

To address these challenges, we propose a teacher-student framework for speaker-invariant spoofing detection that disentangles speaker information while preserving spoofing-relevant cues. An auxiliary teacher model is trained for speaker discrimination using large-scale speaker recognition data~\cite{voxceleb2}, enabling it to encode speaker information in its embeddings. In both the teacher and the student, SSL representations are aggregated with multi-head factorized attentive pooling (MHFA)~\cite{MHFA_ASV}. The teacher then guides the student spoofing detection model through a Gradient Reversal Layer (GRL)~\cite{pmlr-2015-GRL}, encouraging the student to learn representations that are invariant to speaker cues while remaining discriminative for spoofing detection.
While teacher embeddings encode speaker information, they are often entangled with latent features learned during training.
% Adversarial suppression alone can be overly aggressive: the student may remove not only speaker information but also other correlated cues that are beneficial to spoofing detection. 
To prevent the removal of spoofing-related cues, we integrate a Variational Information Bottleneck (VIB)~\cite{alemi2016deepVIB} into the student model. The VIB provides an information-theoretic constraint that regulates the amount of information removed, enforcing a bottleneck where suppression is focused on principal speaker-related information while preserving spoofing-discriminative evidence. This results in a controlled form of invariant learning. Our proposed model achieves a 25.7\% relative reduction to the EER, demonstrating superior generalization across nine evaluation datasets compared to the baseline model.

We make the following contributions.

\begin{itemize}
\item \textbf{Speaker Bias Analysis}: We identify and empirically demonstrate that the distribution of voice identities in the ASVspoof~5 training set creates a spurious correlation between speaker identity and the bona fide/spoof label. We show that models may exploit these voice identity cues as a shortcut, rather than learning the intrinsic discriminative features of spoofing detection.
\item \textbf{Teacher–student speaker-invariant learning}: We introduce a teacher–student framework that suppresses voice identity influences upon spoofing detection without requiring speaker labels for the target dataset.
\item \textbf{Information-controlled adversarial training}: We incorporate a Variational Information Bottleneck to regulate adversarial suppression and control the loss of spoofing-related information.
\item \textbf{Extensive evaluation}: Experimental results demonstrate significant and consistent improvements across nine evaluation datasets.
\end{itemize}

\section{Speaker Bias Analysis}
A primary challenge in synthetic speech detection is the prevalence of speaker bias, where speaker identities are not independent of class labels. When a training distribution exhibits a disjoint speaker set between bona fide and spoofed partitions, classifiers are prone to learning spurious correlations between specific vocal identities and the spoofing label. In such scenarios, the model minimizes empirical risk by identifying who is speaking, leveraging identity-based 'shortcuts', rather than detecting the fundamental acoustic artifacts of the synthesis process. This phenomenon typically stems from two sources. (1) Dataset Design Inconsistency: during data collection, spoofed utterances are often generated without corresponding bona fide references from the same speaker. This creates a distribution shift where the detector learns to distinguish speaker-specific manifolds instead of the spoofing-versus-bona-fide boundary. (2) Acoustic Canonicalization: TTS and VC systems often produce speech with a canonicalized vocal profile. Even when targeting a specific identity, the synthesis process may gravitate toward a "model-average" voice. This results in collapsed speaker variance within the spoofed set compared to the natural diversity of the bona fide set, allowing the detector to rely on these smoothed, deterministic vocal traits. Consequently, models may achieve near-perfect accuracy on intra-dataset evaluations while failing to generalize to unseen speakers and novel attack vectors.

\subsection{Distributional analysis via latent clustering} 

\begin{figure}[t]

    \centering

    \includegraphics[width=1.0\linewidth]{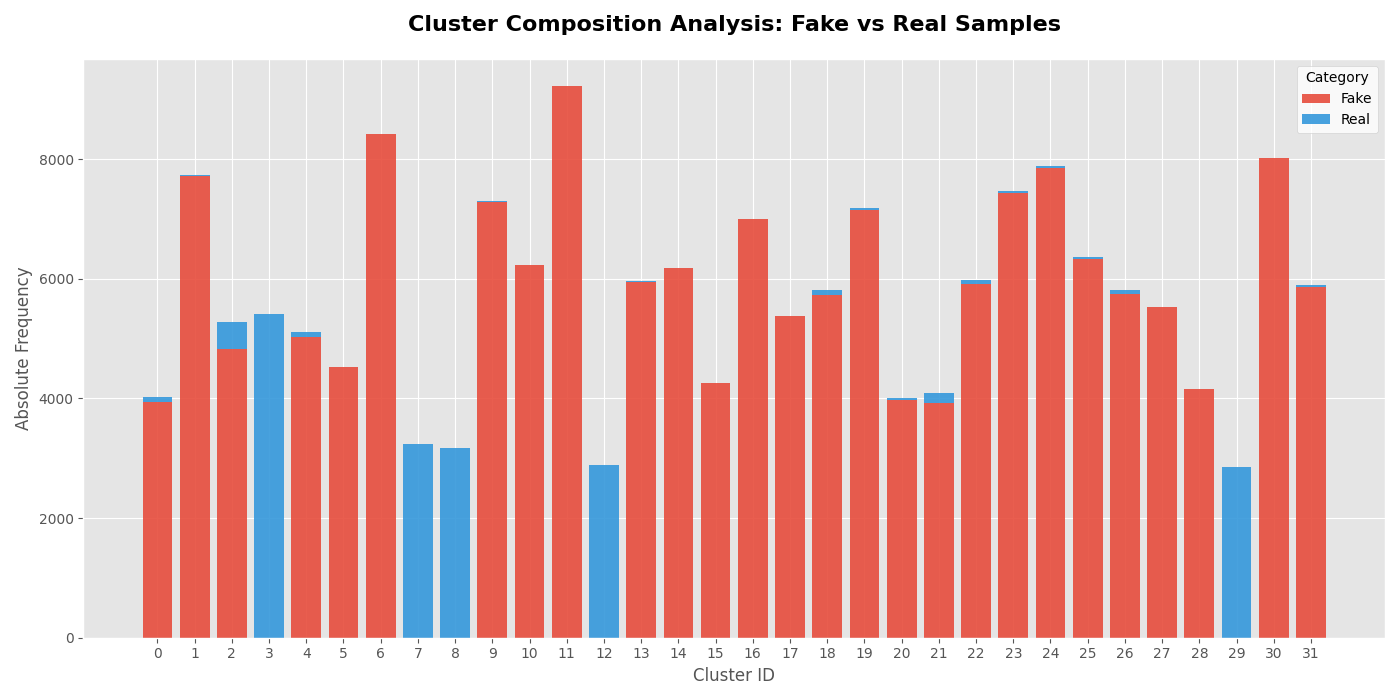}   
    \caption{Cluster composition analysis of {teacher speaker embeddings} for \textbf{ASVspoof 5} training data.}
    \label{fig:cluster1}
\end{figure}

To quantify the extent of speaker bias within the ASVspoof 5 training set, we perform a distribution analysis in the embedding space of a pre-trained speaker recognition model (details in Section \ref{sec:31}). We extract speaker embeddings and partition the manifold using $K$-means clustering ($K=32$). As illustrated in Figure \ref{fig:cluster1}, the resulting cluster assignments reveal a stark structural imbalance and high cluster purity relative to the class labels. Specifically, bona fide utterances are localized within a highly constrained subset of the cluster space, while the vast majority of clusters are comprised exclusively of spoofed utterances. This topological separation confirms a significant speaker-label mismatch, allowing the classifier to reach high training accuracy by identifying speaker-specific clusters rather than underlying generative artifacts.

\begin{figure}[t]

    \centering

    \includegraphics[width=1.0\linewidth]{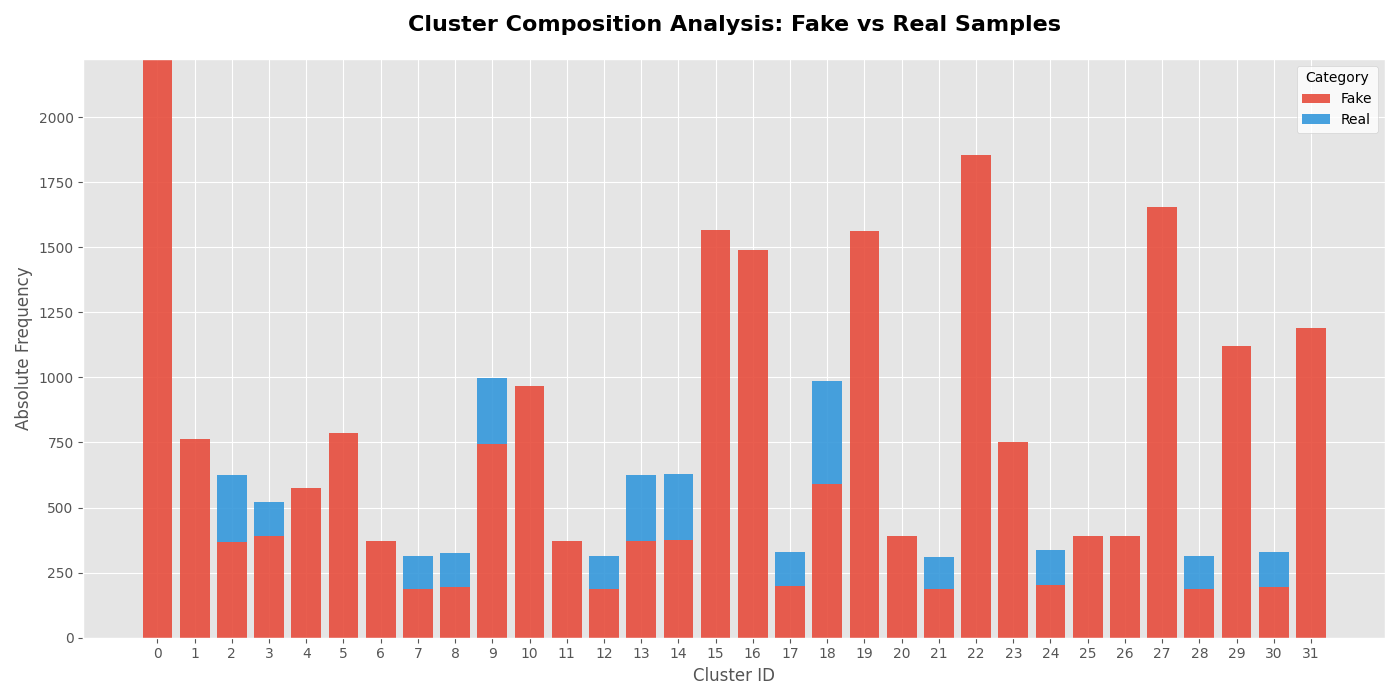}   
    \caption{Cluster composition analysis of {teacher speaker embeddings} for \textbf{ASVspoof 2019} training data.}
    \label{fig:cluster2}
\end{figure}

To benchmark the severity of this bias, we replicate the clustering analysis on the ASVspoof 2019 training set. As demonstrated in Figure \ref{fig:cluster2}, the 2019 dataset exhibits a markedly different topological structure. A substantial proportion of its clusters exhibit high class-entropy, containing a mixture of both bona fide and spoofed utterances. While a subset of clusters remains exclusive to spoofed samples, the degree of identity overlap is significantly higher than in ASVspoof 5. This suggests that ASVspoof 5 presents a heightened risk of identity-based overfitting, necessitating robust regularization or adversarial training to ensure models capture the fundamental characteristics of synthetic speech.

\subsection{Visualizing the impact of acoustic canonicalization} 

\begin{figure}[t]
\label{fig:learned_emb}
    \centering

    \includegraphics[width=1.0\linewidth]{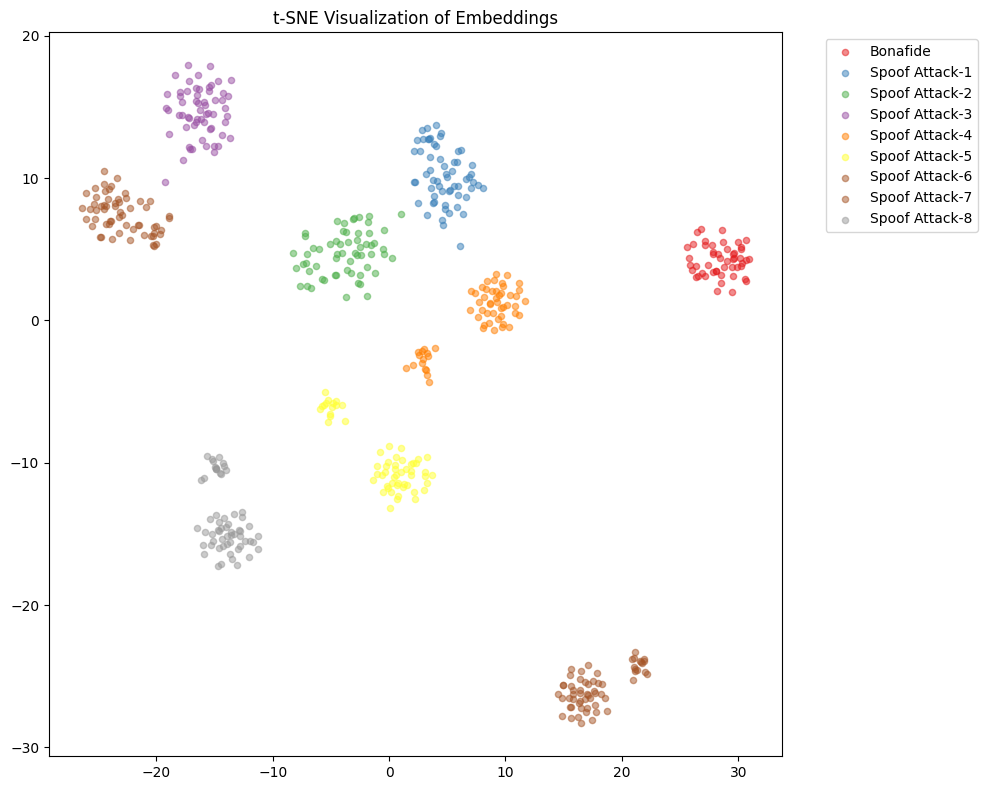}
    
    \caption{t-SNE visualization of {{teacher speaker embeddings}} for a random speaker in \textbf{ASVspoof~5} train.} %Spoofed utterances form distinct clusters based on identical textual content, while bona fide samples reside in a separate manifold, indicating a severe text content mismatch between classes.} 
    \label{fig:learned_emb1}
\end{figure}

\begin{figure}[t]
\label{fig:learned_emb}
    \centering

    \includegraphics[width=1.0\linewidth]{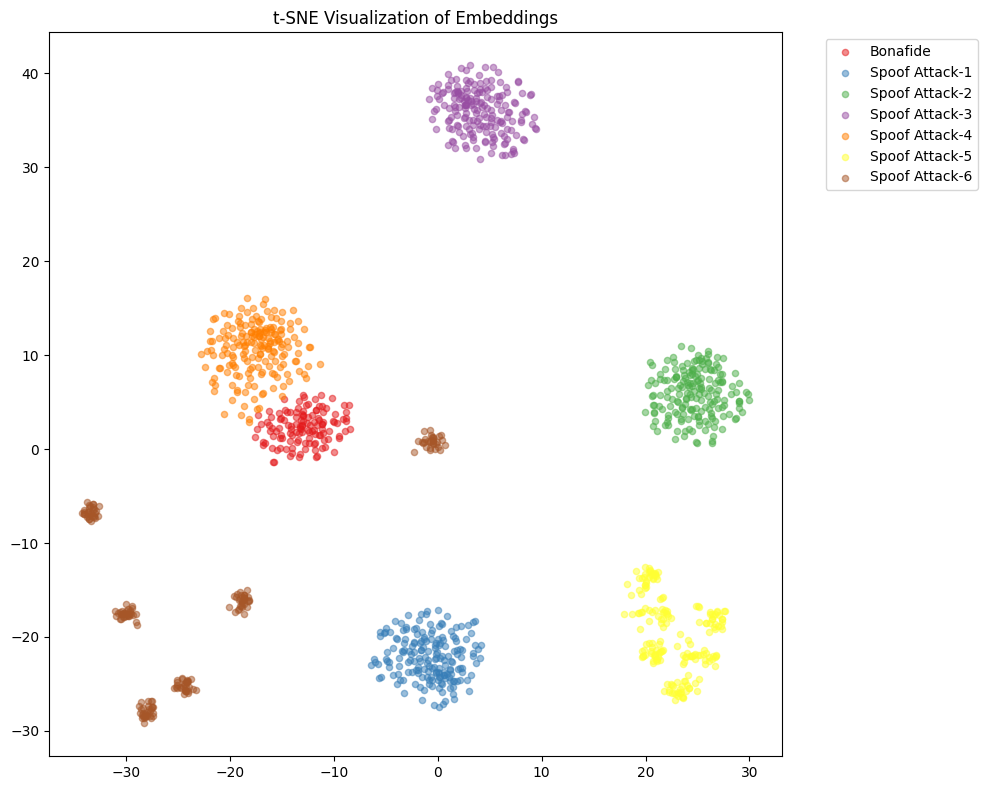}
    
    \caption{t-SNE visualization of {{teacher speaker embeddings}} for a random speaker in \textbf{ASVspoof~2019} train.} %Spoofed utterances form distinct clusters based on identical textual content, while bona fide samples reside in a separate manifold, indicating a severe text content mismatch between classes.} 
    \label{fig:learned_emb2}
\end{figure}

To qualitatively validate the impact of acoustic canonicalization, we perform high-dimensional manifold projection using t-SNE. We randomly select a speaker identity in both ASVspoof 5 and 20T19 and visualize the embeddings of bona fide and spoofed utterances, as shown in Figures \ref{fig:learned_emb1} and \ref{fig:learned_emb2}. 
While the bona fide utterances form a singular cluster, reflecting the natural intra-speaker variance of a human voice, the spoofed samples fragment into distinct sub-clusters corresponding to specific attack algorithms. This topological fragmentation provides empirical evidence of 'model-voice' where each generative system leaves unique and deterministic vocal characteristics. This could create shortcuts of 'model-voice' in model training, thereby compromising the model generalization to unseen attacks.

\section{Proposed Method}

We introduce a teacher-student learning framework augmented with a VIB to suppress {speaker identity information} in spoofing detection.  
The proposed approach enforces speaker invariance at the representation level via GRL guided by a pretrained speaker recognition teacher.

\subsection{Speaker Recognition Teacher Model}
\label{sec:31}

\begin{figure}[t]
\centering
\includegraphics[width=1.0\linewidth]{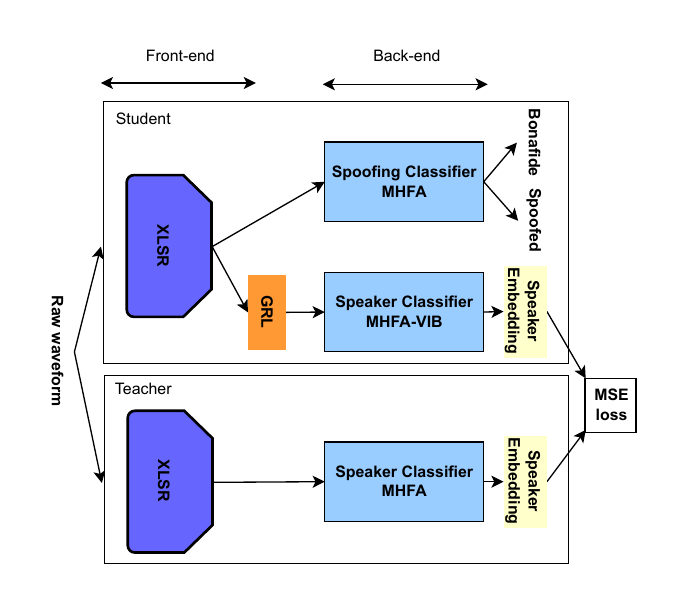}
% (Here you would include a diagram of the architecture)
\caption{The proposed IVSpk-VIB model for speaker-invariant within a teacher-student framework and VIB.}

\label{fig:idfe_arch}
\end{figure}

To model speaker identity information, we first train a speaker recognition teacher model on Voxceleb2 \cite{voxceleb2} dataset which is the most popular for speaker recognition, boasting over 1 million utterances collected from 5,994 speakers.
The architecture consists of a pretrained XLSR encoder followed by a Multi-Head Factorized Attentive (MHFA) backend classifier (Figure \ref{fig:idfe_arch}).  
Once trained, the teacher model is frozen and used solely as an embedding extractor during student training, providing soft labels for adversarial learning.

\subsection{Spoofing Detection Student Model}

The student network is trained to perform spoofing detection while simultaneously suppressing speaker-dependent information.  
As illustrated in Fig.~\ref{fig:idfe_arch}, the student model comprises the following components:

\begin{itemize}
    \item \textbf{Feature Extractor:} A pretrained XLSR encoder~\cite{XLS-R2022} that maps raw audio waveforms into contextualized frame-level representations.
    \item \textbf{Spoofing Classifier Head:} An MHFA-based classifier that predicts the binary spoofing label 
    $\hat{y}_s \in \{\text{bonafide}, \text{spoof}\}$.
    \item \textbf{Speaker Classifier Head:} A MHFA-VIB network dedicated to the speaker classification task. This head is supervised using soft labels from the teacher model's embeddings, with optimization performed via mean squared error (MSE) loss between the student and teacher embeddings.    
    \item \textbf{Gradient Reversal Layer (GRL):} Inserted between the feature extractor and the speaker classifier head. During the forward pass, the GRL functions as an identity transformation. In backpropagation, it scales the gradients by a factor of $-\lambda$ (where $\lambda$ is a hyperparameter), adversarialy encouraging the feature extractor to remove speaker identity information.
\end{itemize}

Through this adversarial learning, the student model is guided to retain spoofing-relevant artifacts while discarding speaker-related information.

\begin{table*}[h]
\renewcommand{\arraystretch}{1.0} % Reduce row height
\centering
\caption{Model performance (EER, \%) across multiple datasets. “Pooled” is obtained by combining all datasets and computing EER with a single global decision threshold, and is used as the main measure of cross-dataset generalization. Bold numbers indicate the best performance.}
\begin{tabular}{lccccccc}
\toprule
& \multicolumn{7}{c}{\textbf{Model (\%EER)}} \\
% \midrule
\cmidrule(lr){2-8}
 % &  &  &  &  &  & MHFA- & MHFA- \\
\textbf{Dataset}  & \textbf{AASIST} & \textbf{Conformer} & \textbf{MHFA} & \textbf{SInMT} \cite{dao2026assessingimpactspeakeridentity} & \textbf{MHFA-VIB} & \textbf{MHFA-IVSpk} & \textbf{MHFA-IVSpk-VIB} \\
\midrule
{\textbf{ITW}}        &7.03   & 5.68  & 4.30 & 3.58 & 3.95  & {2.37}  & \textbf{2.31}\\
{\textbf{ASV 19 eval}} &10.79  & 10.80 & 9.48 & {6.30} & 7.24  & {6.61}  & \textbf{5.18}\\
{\textbf{ASV 21 LA}}    &11.99  & 10.93 & 11.55& 7.35 & 9.41  & {7.15}  & \textbf{6.12}\\
{\textbf{ASV 21 DF}}    &5.29   & 5.54  & 4.83 & {4.09} & 5.49  & {4.15}  & \textbf{3.71}\\
% {ASV5 eval}  &4.10   & 2.86  & 3.76 & 4.01 & 12.07 & 4.73  & 5.19\\
{\textbf{FoR}}        &5.65   & 10.60 & 11.52 & 10.07 & 6.14  & \textbf{5.16}  & {5.38}\\
{\textbf{CodecFake}}  &38.67  & 30.30 & 30.33 & 29.78 & 24.69 & {24.00} & \textbf{22.07}\\
{\textbf{DFADD}}      &10.03  & 5.82  & {2.11} & 2.30 & {2.80}  & \textbf{1.72}  & {4.13}\\
{\textbf{LibriSeVox}} &23.18  & 22.83 & 7.82  & 8.48 & {2.99}  & 5.26  & \textbf{2.62}\\
{\textbf{SONAR}}      &19.12  & 22.60 & 24.37 & 20.97& \textbf{9.37}  & {21.32} & {23.17}\\
% {Average}    &14.64  & 13.90 & 11.81 & 10.32 & \textbf{8.01}  & {8.64}  & {8.30}\\
\midrule
{\textbf{Pool}}       &19.98  & 15.58 & 13.67 & 12.83 & 11.83 & {11.37}  & \textbf{10.15}\\

% \textbf{Average}    &13.58  & 12.80 & 11.01 & 8.01  & \underline{7.44}  & \textbf{6.17}\\
% \textbf{Pool}       &13.75  & 11.41 & 11.94 & 11.83 & \underline{9.90}  & \textbf{9.25}\\

\bottomrule
\end{tabular}
\label{table1}
\end{table*}

\subsection{VIB for Speaker Invariance}
We introduce a VIB regularization into the speaker classifier branch of the student model to constrain the amount of information that can flow during adversarial training, ensuring that suppression is focused on dominant speaker-related information while preserving spoofing-discriminative evidence. This combination enables controlled and principled invariant learning. Given multi-layer XLSR representations 
$\mathbf{o} \in \mathbb{R}^{D \times T \times L}$, where $L$
denotes the number of transformer layers, MHFA first learns
layer-wise importance weights for the key and value streams.
After softmax normalization, the weighted sum across layers is computed as
\begin{equation}
\mathbf{k} = \sum_{l=1}^{L} w_k^{(l)} \mathbf{o}^{(l)}, \quad
\mathbf{v} = \sum_{l=1}^{L} w_v^{(l)} \mathbf{o}^{(l)},
\end{equation}
followed by a linear compression to obtain
$\mathbf{k}, \mathbf{v} \in \mathbb{R}^{d_c \times T }$.

To impose an information bottleneck, the VIB module is applied
to the key representation $\mathbf{k}$.
By bottlenecking $\mathbf{k}$, we prevent the attention mechanism from attending to noisy or redundant features within the high-dimensional SSL embedding. Unlike applying VIB to the final output of MHFA which has already collapsed the sequence into a static embedding, regularizing $\mathbf{k}$ ensures that the aggregation process itself is robust to spurious correlations, leading to a more discriminative feature representation for the backend classifier.
Specifically, a stochastic latent variable $\mathbf{z}$ is modeled by a
Gaussian posterior:
\begin{equation}
q_\theta(\mathbf{z} \mid \mathbf{k}) =
\mathcal{N}(\boldsymbol{\mu}(\mathbf{k}),
\mathrm{diag}(\boldsymbol{\sigma}^2(\mathbf{k}))),
\end{equation}
where the mean $\boldsymbol{\mu}$ and log-variance
$\log \boldsymbol{\sigma}^2$ are predicted by neural networks.
Sampling is performed using the reparameterization trick:
\begin{equation}
\mathbf{z} = \boldsymbol{\mu} + \boldsymbol{\epsilon} \odot
\boldsymbol{\sigma}, \quad
\boldsymbol{\epsilon} \sim \mathcal{N}(\mathbf{0}, \mathbf{I}).
\end{equation}

The variational information bottleneck is enforced through a
Kullback-Leibler (KL) divergence regularization term, defined as
\begin{equation}
\mathcal{L}_{\text{VIB}} =
\mathrm{KL}(q_\theta(\mathbf{z} \mid \mathbf{k}) \,\mid\mid\, r(\mathbf{z})),
\end{equation}
where $q_\theta(\mathbf{z} \mid \mathbf{k})$ denotes the variational
posterior over the latent representation and
$r(\mathbf{z}) = \mathcal{N}(\mathbf{0}, \mathbf{I})$ is a standard normal
prior.
% This term penalizes deviations of the posterior from the prior,
% thereby constraining the information flow from $\mathbf{k}$ to
% $\mathbf{z}$ and enforcing a compact latent representation.
This term acts as an information bottleneck, regularizing the latent space by minimizing the mutual information between $\mathbf{k}$ and $\mathbf{z}$ to ensure a highly compressed representation.

% The latent representation $\mathbf{z}$ is projected to attention logits
% and normalized across the temporal dimension to obtain attention weights.
% These weights are applied to the corresponding value
% representations $\mathbf{v}$, producing weighted value vectors at each
% time step. The weighted values are then aggregated over time to form a compact
% utterance-level embedding for the speaker sub-task.
The latent representation $\mathbf{z}$ is transformed into temporal attention weights, which guide the weighted aggregation of value representations $\mathbf{v}$. This process yields an utterance-level embedding tailored for the speaker sub-task.

\subsection{Speaker-Invariant Teacher-Student Training}

Let $f(\cdot)$ denote the student feature extractor, $g_s(\cdot)$ the spoofing classifier, $g_{d}(\cdot)$ the speaker classifier head, and $g_t(\cdot)$ the frozen speaker teacher.

Given an input utterance $x$, the overall optimization objective is:
\begin{equation}
\min_{f, g_s, g_{d}}
\mathcal{L}_s(g_s(f(x)), y_s)
+ \alpha \mathcal{L}_{d}(g_{d}(\mathrm{GRL}(f(x))),\, g_t(x))
+ \beta \mathcal{L}_{\mathrm{VIB}},
\label{eq:total_loss}
\end{equation}
where:
\begin{itemize}
    \item $\mathcal{L}_s$ is the cross-entropy loss for spoofing detection,
    \item $\mathcal{L}_{d}$ is the mean squared error between student and teacher speaker embeddings,
    \item $\mathcal{L}_{\mathrm{VIB}}$ enforces the information bottleneck constraint,
    \item $\alpha$ controls the strength of speaker invariance,
    \item $\beta$ controls the degree of information bottleneck.
\end{itemize}

\section{Experimental Setup}

\subsection{Speaker Recognition Dataset}
The speaker recognition teacher model is trained on the VoxCeleb corpus, a widely adopted benchmark in speaker recognition community. VoxCeleb contains over one million speech recordings from 5,994 speakers, covering diverse acoustic conditions and speaking styles.

\subsection{Spoofing Detection Datasets}
All spoofing detection models are trained using the ASVspoof~5 training partition, which includes approximately 180k utterances collected from 400 speakers. To avoid evaluation bias, all experiments are conducted exclusively on out-of-domain datasets. We follow the Speech DF Arena evaluation protocols~\cite{dfarena}, selecting only English-language datasets. This results in nine evaluation datasets: In-the-Wild (ITW)\cite{Wild2022}, ASVspoof 2019~\cite{ASVspoof19}, ASVspoof 2021 LA and DF~\cite{ASVspoof21}, Fake-or-Real (FoR)~\cite{FoR2019}, CodecFake~\cite{codecfake}, DFADD~\cite{dfadd}, LibriSe-Vox~\cite{librisevox}, and SONAR~\cite{SONAR}.
Performance is assessed using the equal error rate (EER). In addition, we report the pooled EER which is computed by merging all evaluation sets and estimating EER using a single global decision threshold. The pooled EER serves as the primary metric for assessing cross-dataset generalization.

\subsection{Data Augmentation}
Standard data augmentation strategies are applied during training using the MUSAN corpus and a real room impulse response (RIR) database~\cite{MUSAN,Reverb2017}. For each training utterance, one of four augmentation techniques is randomly selected:

%, following the procedure described in Section~3.2 of~\cite{dao24_asvspoof}.
\begin{itemize}
    \item \textbf{Reverberation:} Utterances are convolved with real RIRs to simulate reverberant conditions arising from sound propagation in diverse acoustic environments.
    \item \textbf{Speech:} Interfering utterances from different speakers are added to each training sample at signal-to-noise ratios (SNRs) ranging from 13 to 20~dB.
    \item \textbf{Music:} Randomly selected music excerpts from the MUSAN corpus are mixed with each utterance at SNRs between 5 and 15~dB.
    \item \textbf{Noise:} Randomly selected noise segments from the MUSAN corpus are added at SNRs ranging from 0 to 15~dB.
\end{itemize}

\subsection{Implementation Details}\label{sec:implem_details}
During training, input audio is randomly segmented into 4-second crops, whereas full-length utterances are used during evaluation. %Data augmentation is applied using additive noise from the MUSAN corpus and simulated reverberation from a real room impulse response (RIR) database~\cite{MUSAN,Reverb2017}.
All models are optimized with the Adam optimizer~\cite{adam} using a learning rate of $10^{-6}$ and a weighted cross-entropy objective. Training is conducted for 30 epochs with a batch size of 32 on NVIDIA A100 GPUs. The weighting coefficients $\alpha$ and $\beta$ in Eq.~\ref{eq:total_loss} are set to 0.1.
We evaluate three baseline models based on XLSR: AASIST, Conformer, and MHFA. Additionally, we introduce an MHFA variant integrated with a VIB regularization, referred to as MHFA-VIB.
Finally, the SInMT baseline \cite{dao2026assessingimpactspeakeridentity} provides a speaker-invariant benchmark by exploiting speaker labels from ASVspoof 5 training set \cite{dao2026assessingimpactspeakeridentity}.
All baseline models are compared against the proposed speaker-invariant framework with and without the VIB module (i.e. IVSpk and IVSpk-VIB).

\begin{table}[t]
\centering
\caption{Model comparison between our proposed IVSpk-VIB and top-4 submissions to track 1 open condition of ASVspoof 5 challenge.}
\label{tab:evaluation_results}

% \resizebox{\columnwidth}{!}{
% \begin{tabular}{lccccc}
% \toprule
% & \multicolumn{5}{c}{\textbf{Model (\%EER)}} \\
% % \toprule
% \cmidrule(lr){2-6}
%  % &  &  &   &  & {MHFA-} \\
%  Dataset & {T36} & {T27} & {T23} & {T43}  &{MHFA-IVLing-VIB} \\
% \midrule

% {ASV~5 eval}      & 3.37  & \textbf{3.30}  & 4.23  & 4.33 &  {5.19} \\
% {ASV~19 LA}    & 16.27 & 17.33 & 16.73 & 26.63 &  \textbf{5.18} \\
% {ASV~21 LA}    & 15.73 & 18.70 & 13.13 & 25.57 &  \textbf{6.12} \\
% {ASV~21 DF}    & 11.57 & 10.63 & 14.87 & 14.20 &  \textbf{3.71} \\
% {ITW}                 & 14.71 & 13.37 & 10.20 & 6.85  &  \textbf{2.31} \\
% \toprule
% {Average}                 & 12.33 & 12.66 & 11.83 & 15.51 &  \textbf{4.50} \\
% \bottomrule
% \end{tabular}}
% \end{table}

\resizebox{\columnwidth}{!}{
\begin{tabular}{lccccc}
\toprule
 & \multicolumn{5}{c}{\textbf{Model (EER \%)}} \\
\cmidrule(lr){2-6}
\textbf{Dataset} & \textbf{T43} & \textbf{T27} & \textbf{T36} & \textbf{T23} & \textbf{MHFA-IVSpk-VIB} \\
\midrule
ASV~5 eval & 4.33 & \textbf{3.30} & 3.37 & 4.23 & 5.19 \\
ASV~19 LA  & 26.63 & 17.33 & 16.27 & 16.73 & \textbf{5.18} \\
ASV~21 LA  & 25.57 & 18.70 & 15.73 & 13.13 & \textbf{6.12} \\
ASV~21 DF  & 14.20 & 10.63 & 11.57 & 14.87 & \textbf{3.71} \\
ITW        & 6.85  & 13.37 & 14.71 & 10.20 & \textbf{2.31} \\
\midrule
\textbf{Average} & 15.51 & 12.66 & 12.33 & 11.83 & \textbf{4.50} \\
\bottomrule
\end{tabular}}
\end{table}

\section{Results}

\subsection{Baseline Spoofing Detection Performance}

The performance of the baseline architectures (i.e., AASIST, Conformer, and MHFA) varies significantly across the evaluated datasets, underscoring the persistent generalization gap in audio anti-spoofing (see Table 1). While the models perform competently on the ASVspoof 2021 DF task (achieving EERs near 5\%), they struggle with the domain shifts present in more rigorous benchmarks. For example, AASIST reaches an EER of 38\% on CodecFake, while MHFA records 24.3\% on SONAR. MHFA emerges as the most robust baseline, yielding the lowest pooled EER of 13.67\%.

Integrating the VIB module into the MHFA framework (MHFA-VIB) produces performance gains across almost all datasets. The pooled EER improves from 13.67\% to 11.83\%, representing a relative reduction of 13.4\%. The most substantial improvement is observed on the SONAR dataset, where the EER drops from 24.37\% to 9.37\%. This suggests that the VIB objective filters irrelevant features, facilitating the extraction of compact and highly generalizable representations.

\subsection{Performance of Speaker-Invariant Models}

The progression from previous work of standard adversarial training \cite{dao2026assessingimpactspeakeridentity} to our proposed teacher-student framework highlights the critical role of speaker-invariant representation learning. The previous approach, the SInMT model, utilized speaker ID labels from the ASVspoof 5 training set to learn speaker-invariant features for spoofing detection. While SInMT achieved condiderable improvements on datasets like In-the-Wild (ITW) and ASVspoof 2021 (LA/DF), it yielded a modest 6.1\% relative reduction in pooled EER across all nine benchmarks, suggesting that in-domain speaker labels provide insufficient diversity for broad generalization. To address this, we introduced MHFA-IVSpk, a teacher-student framework that leverages robust speaker embeddings from a teacher model pre-trained on the large-scale VoxCeleb dataset. This integration of extensive speaker variance allowed MHFA-IVSpk to outperform the SInMT baseline by 11.3\% in terms of relative pooled EER. The most significant gains were observed in Fake-or-Real (EER dropped from 10.07\% to 5.16\%) and CodecFake (from 29.78\% to 24\%). These results demonstrate that leveraging broad speaker diversity from external sources yields superior cross-domain generalization compared to relying solely on in-domain speaker labels. Finally, the MHFA-IVSpk-VIB model incorporates the VIB to regulate the information flow during adversarial training. This mechanism ensures that the model suppresses the most relevant speaker-related information while preserving the essential spoofing cues. MHFA-IVSpk-VIB achieves a 25.7\% relative improvement in pooled EER compared to the MHFA baseline.

\subsection{Benchmarking Against Top Challenge Submissions}

We evaluate the proposed IVSpk-VIB model against the top four performing systems from the ASVspoof 5 Challenge (Table \ref{tab:evaluation_results}), as detailed in the official organizer report~\cite{wang2026asvspoof5evaluationspoofing}. The results reveal a substantial generalization gap among the top-ranked submissions. For example, while the T27 system achieves a state-of-the-art EER of 3.30\% on the ASVspoof 5 evaluation set, its error rate climbs sharply to 17.33\% on ASVspoof 2019 LA and 18.70\% on ASVspoof 2021 LA.

In contrast, IVSpk-VIB sacrifices some in-domain performance on ASVspoof 5 for enhanced out-of-domain robustness. Compared to the T23 system, IVSpk-VIB achieves a 62\% relative improvement in average EER across all evaluation benchmarks. This performance gain highlights the efficacy of our approach in generalizable spoofing detection.

% \subsection{Performance Analysis on ASVspoof 2021 LA}

\section{Conclusion}

In this paper, we address a critical yet unexplored limitation of spoofing detection systems: speaker bias. We demonstrate that, for the ASVspoof 5 dataset, mismatches between the voice traits of spoofed and bona fide utterances are a potential learning shortcut which can undermine model generalisation. 
To address this weakness, we propose IVSpk-VIB, a teacher–student adversarial framework designed to enforce speaker invariance.
Using a speaker recognition teacher and a gradient reversal layer, our approach suppresses potentially-deceptive voice information without requiring speaker labels. Crucially, the integration of a Variational Information Bottleneck regulates this process, managing the inherent trade-off between the removal of biased voice features and the preservation of essential acoustic cues pertinent to spoofing detection.
Our experimental results reveal two key findings.
(1) The use of a teacher model pre-trained using large-scale speaker recognition data provides a more diverse and effective basis for invariance than limited in-domain speaker labels, yielding an 11.3\% relative improvement in pooled EER.
(2) The integration of a VIB is essential for balancing invariance and discriminability. By regulating adversarial suppression, the VIB protects against the suppression of spoofing cues, leading to a 10.7\% relative gain in pooled EER compared to the same system without VIB and a 25.7\% gain relative to the MHFA baseline.

% In this work, we mitigate the critical challenge of speaker-induced shortcut learning in spoofing detection. We demonstrate that while modern detectors achieve high in-domain performance, they often fail to generalize because they implicitly associate speaker characteristics with spoofing labels. To mitigate this, we introduce a teacher-student adversarial framework that leverages external speaker knowledge from the VoxCeleb dataset, thereby bypassing the requirement of speaker labels for spoofing datasets.
% Our experimental results validate two key findings.
% (1) Utilizing a teacher model pre-trained on large-scale speaker recognition data provides a more diverse and effective basis for invariance than limited in-domain speaker labels, yielding an 11.3\% relative improvement over prior adversarial methods.
% (2) The integration of a VIB is essential for balancing invariance and discriminability. By regulating adversarial suppression, the VIB prevents the "over-erasure" of beneficial cues, leading to a relative gain of 25.7\% in pooled EER over the MHFA baseline.

\section{Acknowledgements}

This work was performed using HPC resources from GENCI-IDRIS. This work was financially supported by ANR BRUEL (ANR-22-CE39-0009).

\bibliographystyle{IEEEtran}
\bibliography{Odyssey2026_BibEntries}

\end{document}